\documentclass[conference]{IEEEtran}
\IEEEoverridecommandlockouts
\usepackage{cite}
\usepackage{amsmath,amssymb,amsfonts}
\usepackage{graphicx}
\usepackage{textcomp}
\usepackage{xcolor}
\usepackage{xspace}
\usepackage{cprotect}
\def\BibTeX{{\rm B\kern-.05em{\sc i\kern-.025em b}\kern-.08em
    T\kern-.1667em\lower.7ex\hbox{E}\kern-.125emX}}

\begin{document}


\newcommand{\vect}[1]{\boldsymbol{\mathrm{#1}}}
\newcommand{\vectt}[1]{\boldsymbol{\mathrm{#1}}_t}
\newcommand{\vectall}[1]{\vect{#1}_{1:T}}

\def\remRF#1{{\noindent\color{red}{{\footnotesize [RF: #1]}}}}
\def\addRF#1{{\noindent\color{red}{#1}}}

\newcommand{\rf}[1]{\textcolor{red}{[RF: #1]}}
\newcommand{\dn}[1]{\textcolor{blue}{[DN: #1]}}
\newcommand{\ms}[1]{\textcolor{orange}{[MS: #1]}}
\newcommand{\rv}[1]{\textcolor{green}{[RV: #1]}}
\newcommand{\ct}[1]{\textcolor{orange}{[CT: #1]}}
\newcommand{\todo}[1]{\textcolor{blue}{[TODO: #1]}}
\newcommand{\revise}[1]{#1}

\newcommand{\GeoTrackNet}{\textit{GeoTrackNet }}


\title{Detection of Abnormal Vessel Behaviours from \\AIS data using GeoTrackNet: \\
from the Laboratory to the Ocean}

\author{Duong Nguyen$^1$, Matthieu Simonin$^2$, Guillaume Hajduch$^3$, Rodolphe Vadaine$^3$, C\'edric Tedeschi$^2$, Ronan Fablet$^1$\\
(1) IMT Atlantique, Lab-STICC, 29238 Brest, France\\
(2) University of Rennes, Inria, CNRS, IRISA\\
(3) CLS: Collecte Localisation Satellites, 29280 Brest, France}


\maketitle

\begin{abstract}

The constant growth of maritime traffic leads to the need of automatic anomaly detection, which has been attracting great research attention. Information provided by AIS (Automatic Identification System) data, together with recent outstanding progresses of deep learning, make vessel monitoring using neural networks (NNs) a very promising approach. This paper analyses a novel neural network we have recently introduced ---\textit{GeoTrackNet}--- regarding operational contexts. 
Especially, we aim to evaluate (i) the relevance of the abnormal behaviours detected by \textit{GeoTrackNet} with respect to expert interpretations, (ii) the extent to which \textit{GeoTrackNet} may process AIS data streams in real time.  We report experiments showing the high potential to meet operational levels of the model.
\end{abstract}

\section{Context}
\label{sec:context}

Maritime traffic surveillance has played an important role, not only in navigation  safety, but also in other Maritime Situational Awareness (MSA) aspects. Among others, the early detection of abnormal behaviours helps identify suspicious activities, enforce law, perform efficiently search and rescue, etc. 

In recent years, the world has experienced an unprecedented development of Big Data and Artificial Intelligence (AI), of which Deep Learning~\cite{lecun_deep_2015} is the core. People have tried to leverage Big Data and AI in many domains, and maritime surveillance is not an exception. The rich information provided by the Automatic Identification System (AIS) makes it a very appealing research topic. In~\cite{mantecon_deep_2019}, Airbus Defence and Space used a ResNet-based architecture to re-identify the navigation status of vessels. Similarly, convolutional neural networks were used to discover mobility modes in~\cite{chen_use_2018}. Some work went into detecting fishing pattern using a data mining approach~\cite{souza_improving_2016}. NATO STO Centre for Maritime Research and Experimentation used recurrent neural networks to predict vessel position in~\cite{forti_prediction_nodate}. Our previous work~\cite{nguyen_multi-task_2018,nguyen_geotracknet-maritime_2019} takes advantage of the high modeling capacity of deep neural network to create a probabilistic presentation of AIS tracks for anomaly detection. 

Although those models have given impressive experimental results, applying them to operational systems is still called into question: (i) the black (or gray) box nature of Deep Learning raises concern about their reliability, (ii) the massive volume of AIS data requires a suitable deployment strategy. Besides that, the absence of reference groudtruthed dataset also questions the ability to evaluate the operational usefulness of such learning-based strategies.

In this paper, we report our progress in making a research prototype, namely the \GeoTrackNet model presented in \cite{nguyen_multi-task_2018} and \cite{nguyen_geotracknet-maritime_2019}, reach operational needs. 
\revise{Specifically, we analyse the types of anomalies detected by the model, and evaluate scalability as well as the possibility of deploy it in real-time a big data platform.}

The paper is organised as follows: in Section~\ref{sec:geoTrackNet}, we give a brief introduction of \textit{GeoTrackNet}. The report on the validation of the output of the model is presented in Section~\ref{sec:validation}. Section~\ref{sec:onlineDetection} demonstrates the ability to deploy \GeoTrackNet in a real-time distributed system. We end the paper by the conclusions and perspectives in Section~\ref{sec:conclusions}.
\section{GeoTrackNet}
\label{sec:geoTrackNet}

In this section, we summary the principles of \textit{GeoTrackNet}. For more details, readers are encouraged to refer to~\cite{nguyen_geotracknet-maritime_2019}.

\begin{figure}
    \centering
    \includegraphics[width=85mm]{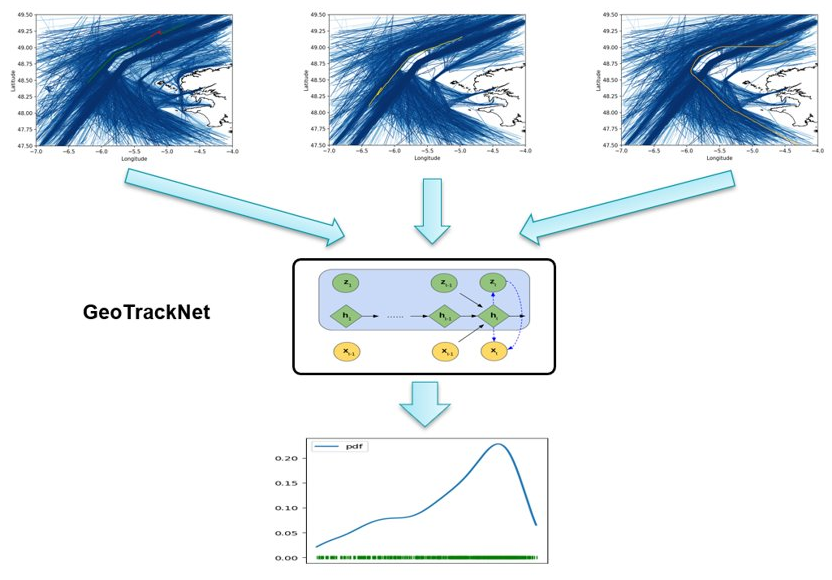}
    \caption{\GeoTrackNet principle. In the training phase, the model learns a distribution representing AIS tracks in the ROI. Any track that does not follow this distribution will be associated with a very low likelihood and will be considered as abnormal in the test phase.} 
    \label{fig:GeoTrackNet}
\end{figure}

\GeoTrackNet is a probabilistic model to detect abnormal behaviours in maritime traffic. The model is based on the hypothesis that most of vessels' behaviours in the training set are normal and the model can learn the dynamics of those tracks. In the training phase, \GeoTrackNet learns a distribution representing the data (see Fig.~\ref{fig:GeoTrackNet}). This distribution is then used in the test phase to state how likely a new AIS track is. 
Any track that does not follow this distribution will be flagged as abnormal. To capture the high complexity of AIS data, \GeoTrackNet uses a Variational Recurrent Neural Network (VRNN)~\cite{chung_recurrent_2015}, which is the state of the art in text and speech modeling. 

Because vessel density and vessels' behaviours are not geographically homogeneous, \GeoTrackNet also exploits a geographically-dependent \textit{a contrario} detection rule. The Region Of Interest (ROI) is divided into small cells, then a local threshold is applied to state the normality of each AIS message in this cell. 

\section{Analysis of the performance of GeoTrackNet}
\label{sec:validation}

Bringing such research prototype to an operational system requires, \textit{inter alia}, to address its validation in terms of relevance of the unusual trajectories that are detected, the explainability of the results and the scalability of the model. 

\subsection{Relevance of the detected anomalies}
\label{sec:analysis_relevance}

Since \GeoTrackNet is a probabilistic model, it detects what is probabilistically unusual. These anomalies may not correspond to suspicious activities. As mentioned above, maritime anomaly detection is an intermediate step in MSA. The final purpose is to detect whether actions are needed when an anomaly happens, e.g. enforcing law when a smuggling activity is committed. It is thus operationally important to better understand the types of unusual trajectories raised by the system together with its limitations\footnote{\revise{The aim of this paper is to analyse the types of vessels' behaviours flagged as abnormal by \textit{GeoTrackNet}. For the comparison of the performance of of \textit{GeoTrackNet} and other state-of-the-art models', please refer to \cite{nguyen_geotracknet-maritime_2019}}}.

\revise{It is also important to note that maritime traffic anomaly detection is an ill-defined problems. No universal definitions of abnormal behaviours or groundtruthed datasets are available for this problem, hence the validation of the outcomes of the detector is subjective. We focus on understanding the types of vessels' behaviours will be flagged by \textit{GeoTrackNet}, rather than calculating quantitative measure.}

This analysis involved a manual inspection of the results of \GeoTrackNet over AIS data acquired around the Ushant Traffic Separation Scheme (TSS) by maritime domain experts
from CLS Group  (Collecte Localisation Satellites). 
Those experts are in charge of the operational monitoring of maritime traffic using multiple sources of data (AIS, Earth Observation, Open Source Intelligence, ELINT, etc). For this analysis they took into consideration their knowledge of the maritime activities in the area, the AIS tracks, the log of the vessels, the weather conditions, etc. This analysis allows to highlight the types of trajectories that are raised (or not raised) as unusual.

\subsubsection{Unusual trip}
\label{sec:unusual_trip}

\begin{figure}
    \centering
    \includegraphics[width=80mm]{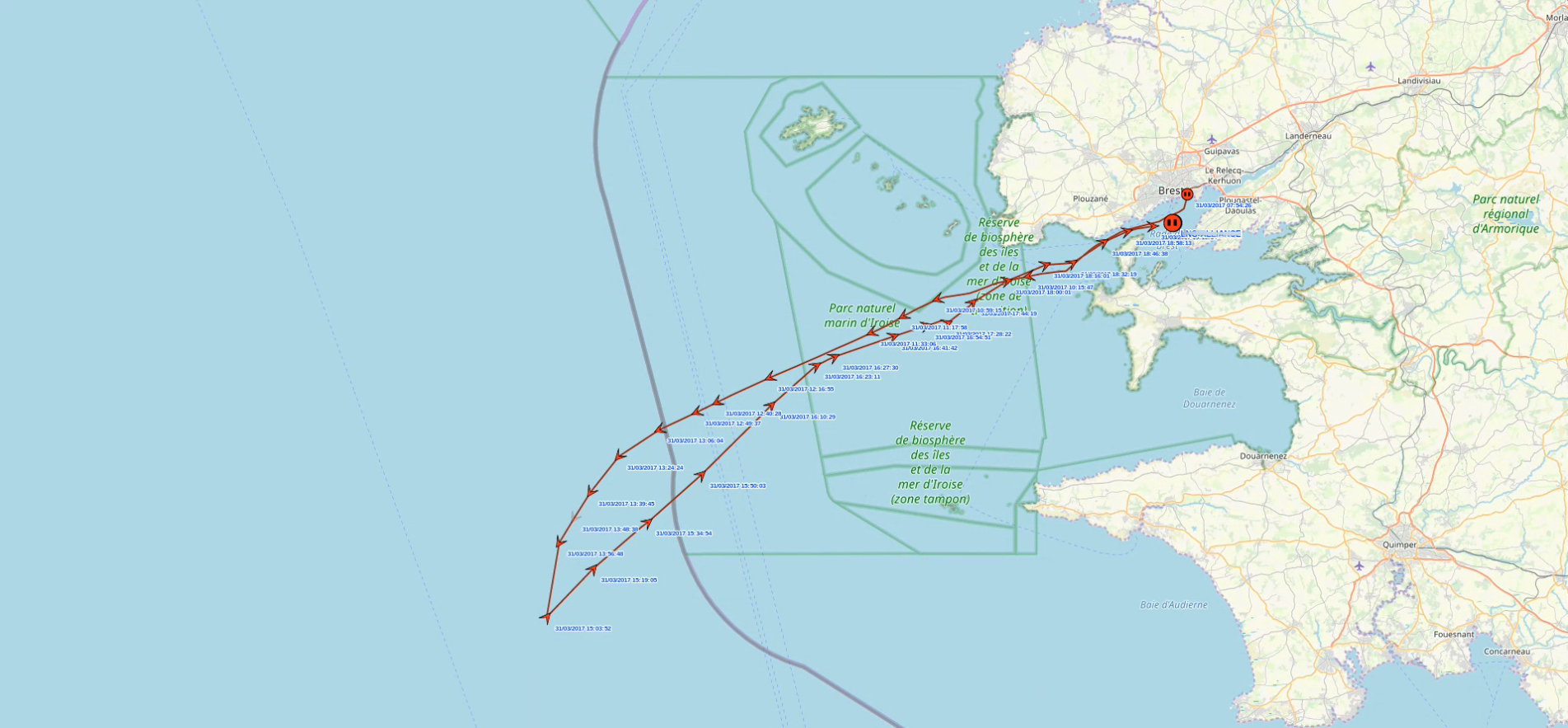}
    \caption{Running-in. This vessel made a test voyage after being repaired at the port of Brest.}
    \label{fig:running-in}
\end{figure}

Fig.~\ref{fig:running-in} shows the trajectory of a vessel of type cargo raised by \GeoTrackNet, steaming to sea and then turning back to the departing port. The visual inspection of the vessels tracks in the dataset shows that this behaviour is indeed unusual. One could consider that this voyage might involve a transshipment. However, after checking other sources of information, we figured out that this vessel had been repaired in the port of Brest (it was berthed in the shipyard), and most probably performed a test trip before going back to the port.
Even if finally not suspicious, we consider relevant to report such tracks as statistically unusual in this location.

\subsubsection{Effect of weather conditions}
\label{sec:weather_conditions}

\begin{figure}
    \centering
    \includegraphics[width=80mm]{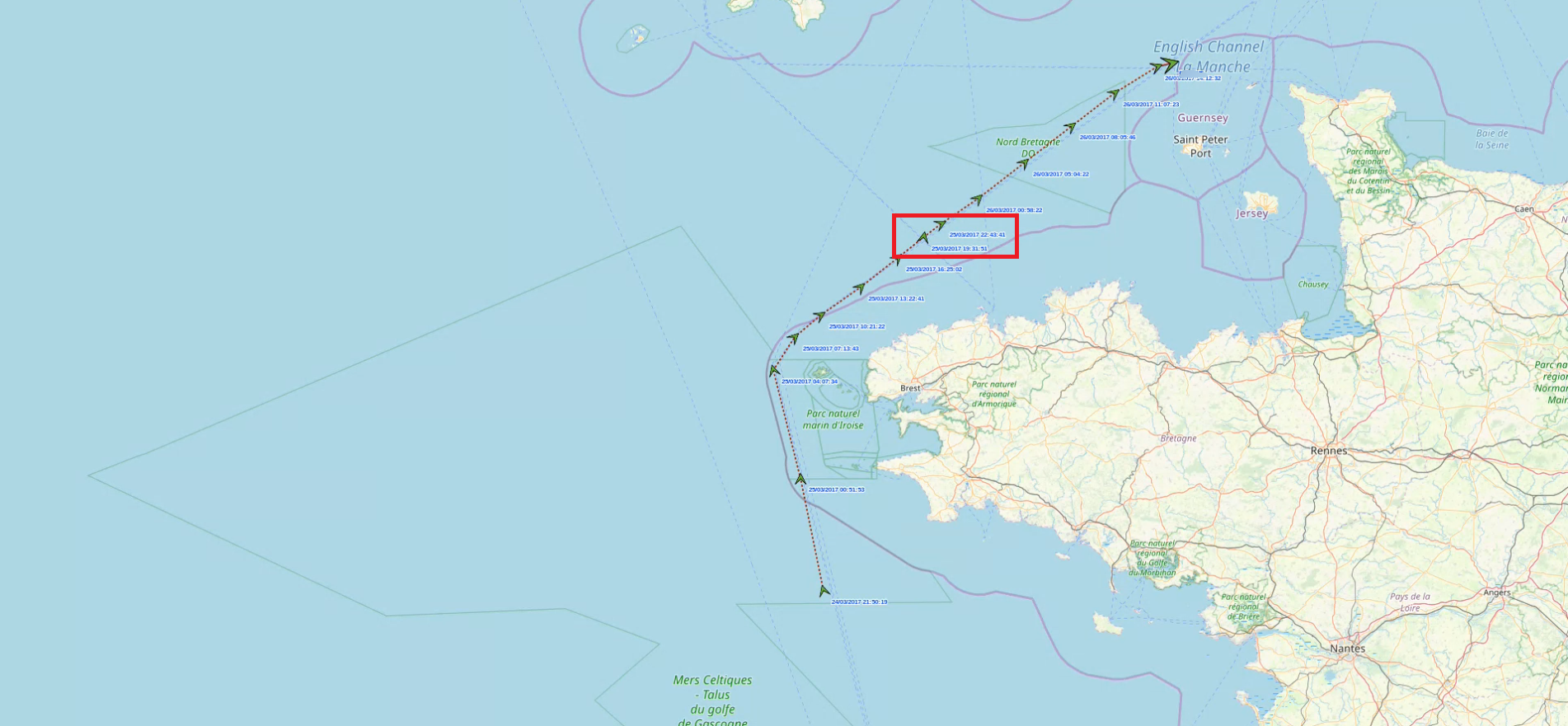}
    \caption{Anomaly due to extreme weather condition. Because of a strong wind blew from the opposite direction, this vessel slowed down (in the red zone). Some similar cases were detected at the same night.}
    \label{fig:wind}
\end{figure}

\begin{figure}
    \centering
    \includegraphics[width=80mm]{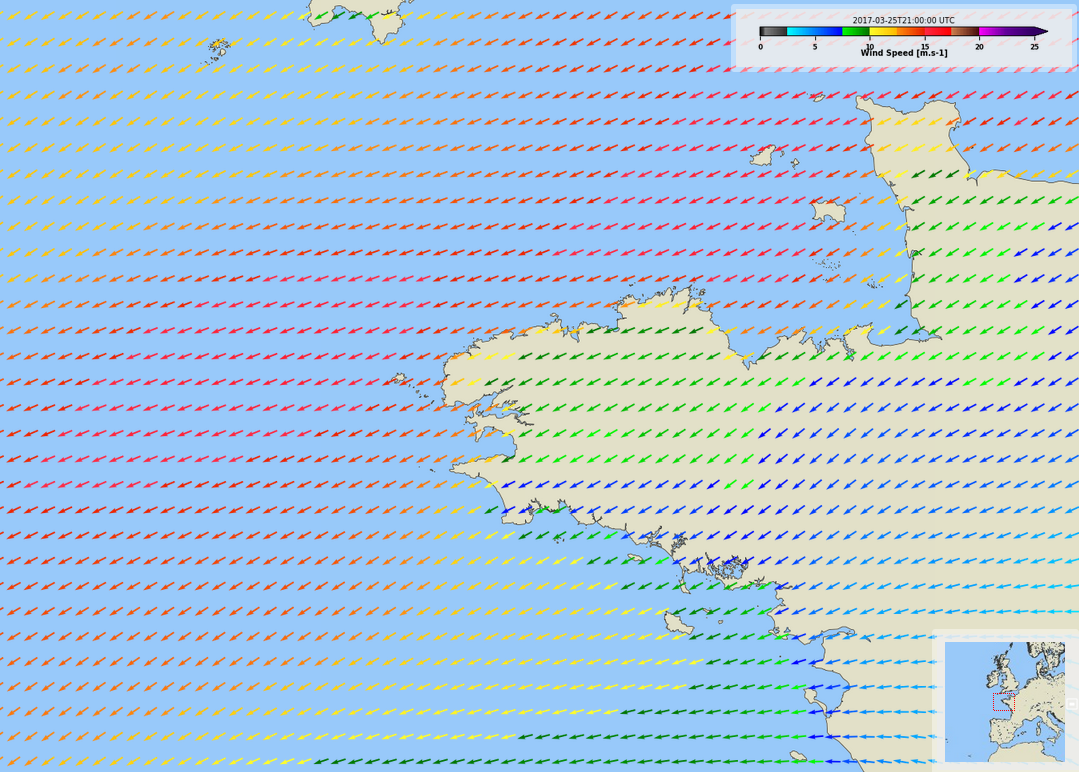}
    \caption{Wind speed and wind direction on 2017-03-25 at 21:00 UTC from NCEP GFS weather forecast. From this model the wind was blowing from North East in the opposite direction of vessels entering the English Channel with a speed of around 30 knots.}
    \label{fig:wind_speed}
\end{figure}

\GeoTrackNet does not take into account the environmental conditions.
In extreme weather/ocean conditions, vessels' behaviours change.
By nature, extreme conditions are rare events. Hence, the model did not see enough of these types of behaviours in the training set to include them into the normalcy model. 
The changes of behaviour due to specific weather condition may then be flagged  as abnormal behaviours. 

An example of such cases is shown in Fig.~\ref{fig:wind}. On the night of March 25, 2017, there was a strong wind blowing from the North East. Some vessels in the area were highly affected by this wind at the entrance of the English Channel just after turning right after the Ushant TSS, while the wind was blowing in the opposite direction with a speed of around of 30 knots.

Again, this behaviour is not suspicious, but raised as statistically unusual. We consider relevant to report such tracks as in such specific weather conditions. Some vessels may be impacted differently by wind drag.

\subsubsection{Weak changes of trajectories}
\label{sec:weak}

\begin{figure}
    \centering
    \includegraphics[width=80mm]{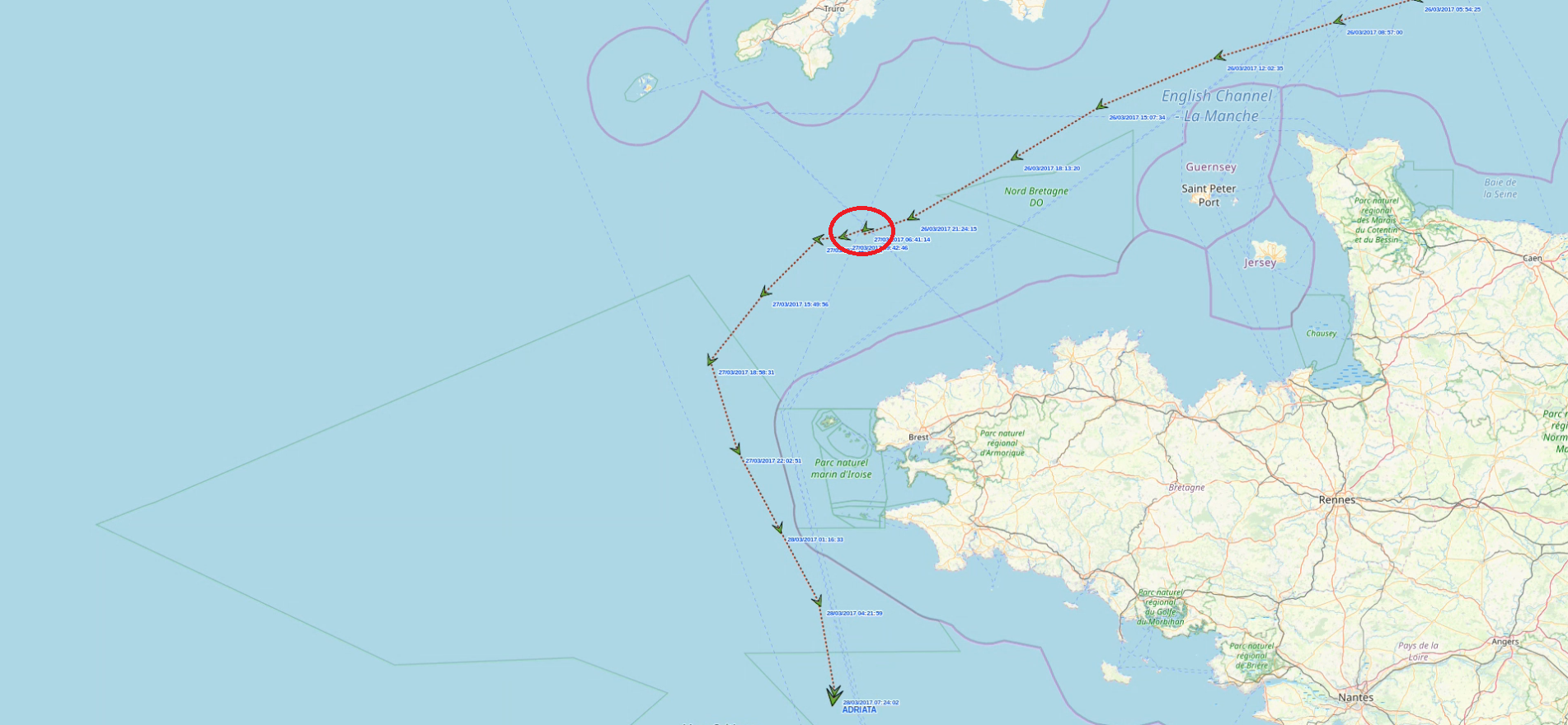}
    \caption{Example of a weak change in trajectory detected by \GeoTrackNet.}
    \label{fig:stop}
\end{figure}

As expected \GeoTrackNet succeeds to raise anomalies on small changes of trajectories mixing both unusual changes of speed and location. Such weak change of trajectory is illustrated in Fig.~\ref{fig:stop}. A vessel left the English channel, slowed down slightly off the usual route, and then started again. Similar behaviours were observed near the TSS in the opposite track (vessel stopping briefly before entering the Channel).

We associated this type of behaviours to the \textit{MARPOL NOx Tier III Regulation before entering/leaving Sulphur Emission Control Areas (SECA)}~\cite{international_maritime_organization_imo_international_nodate}. 
Such SECA has been in place since 2015 for the North Sea including the English Channel with a limit at the 5°W meridian. The reporting of such behaviours is relevant as it is likely that the vessels stop their engine before changing their fuel. This may relate to  higher risks of engine failure (not restart after stop) with vessels potentially drifting in the traffic lane or to the coast.

\subsubsection{Unreported anomalies}
\label{sec:unreported}

As \GeoTrackNet is designed to report statistically unusual tracks, it does not report those that are not sufficiently represented in the training dataset. Without surprise, it thus fails to report vessels crossing the traffic lanes and entering the separation area. This TSS is heavily monitored by the local authorities and the Maritime Rescue Co-ordination Centre (MRCC) and the separation areas can easily be seen for instance in Fig. \ref{fig:new_results}. However, such trajectories can easily be reported by simpler techniques, by just checking the position of AIS messages with respect to those forbidden areas.

\subsection{Scalability of the model}
\label{sec:scalability}

Former results were demonstrated on an area of interest with a limited geographic coverage based on training over terrestrial AIS data~\cite{nguyen_geotracknet-maritime_2019}. Using \GeoTrackNet on different scenarios raises the following questions: Does it provide similar results when trained and applied over a larger geographic area? Does it provide similar performance using a combination of both terrestrial and satellite AIS data? Does it provide relevant results on areas where maritime traffic is less structured? The evaluation of these questions is ongoing. However, we have preliminary results related to the scalability of the model.

\begin{figure}
    \centering
    \includegraphics[width=80mm]{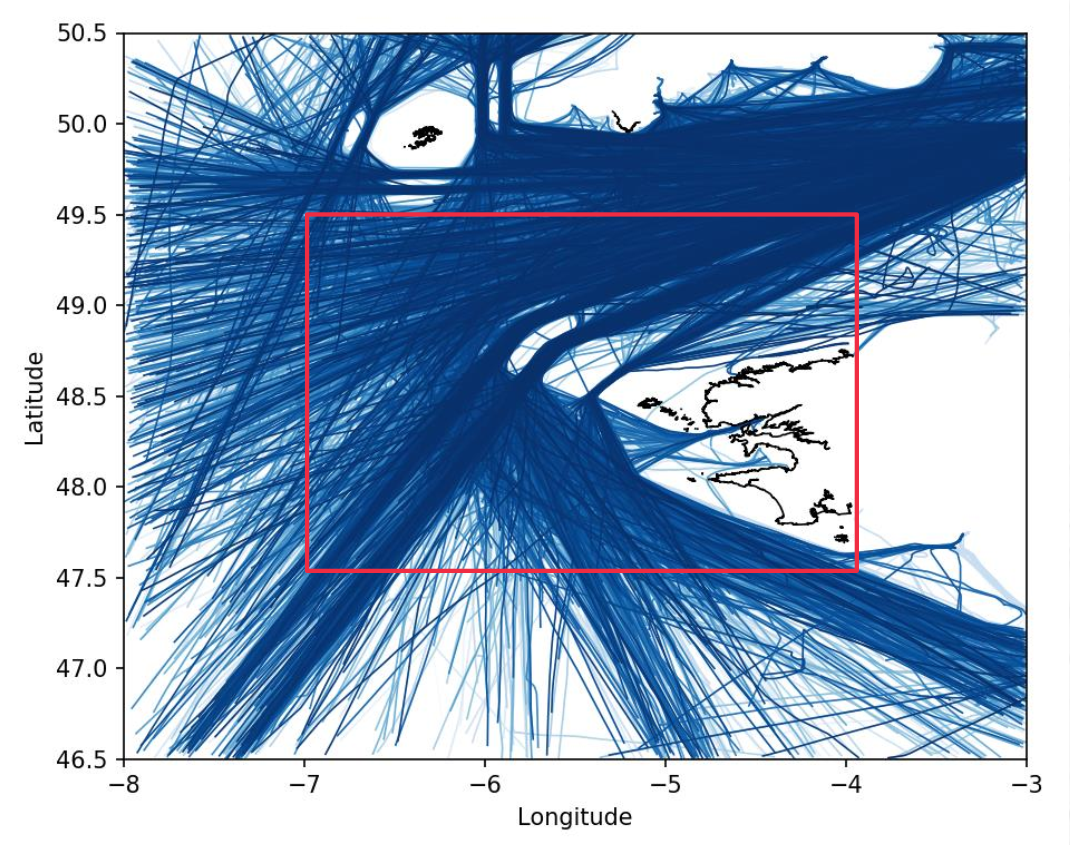}
    \caption{New dataset. The ROI of the new dataset is 3.4 times bigger than the original. It contains both S-AIS and T-AIS, while the original training set contains T-AIS only.}
    \label{fig:new_ROI}
\end{figure}

We tested the model on a new dataset, which contains both satellite AIS messages (S-AIS) and terrestrial AIS messages (T-AIS)\footnote{The original dataset contains T-AIS only.}, over the same period (January to March 2017). The ROI of the new dataset is 3.4 times bigger than the original one. We used the same hyper-parameters as those used in the previous setting, except the dimensions of the ROI. 

\begin{figure}
    \centering
    \includegraphics[width=80mm]{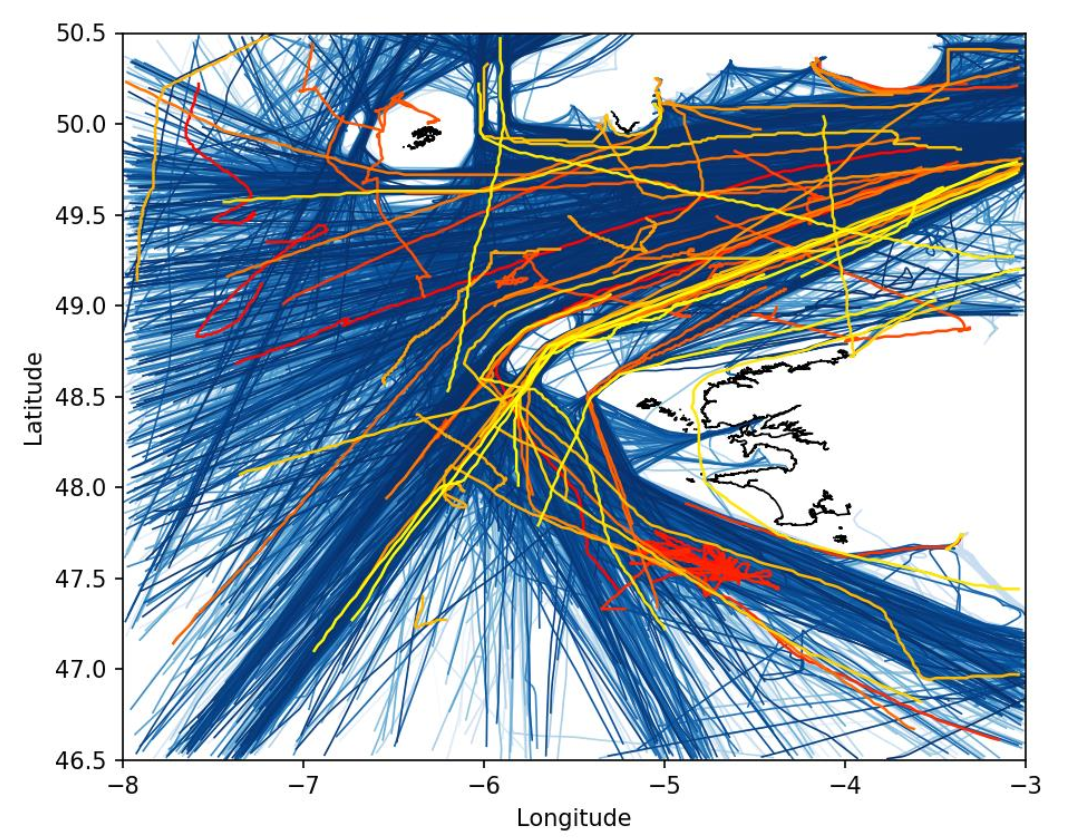}
    \caption{AIS tracks detected as abnormal by \GeoTrackNet trained on the new dataset. Blue: AIS tracks in the training set; other colours: detected tracks.}
    \label{fig:new_results}
\end{figure}

The result is shown in Fig.~\ref{fig:new_results}. Almost all the tracks flagged as abnormal by the previous model are detected again by the new model. There are new abnormal tracks because they do not exist in the original dataset (S-AIS tracks or outside of the previous ROI). 

\begin{figure}
    \centering
    \includegraphics[width=80mm]{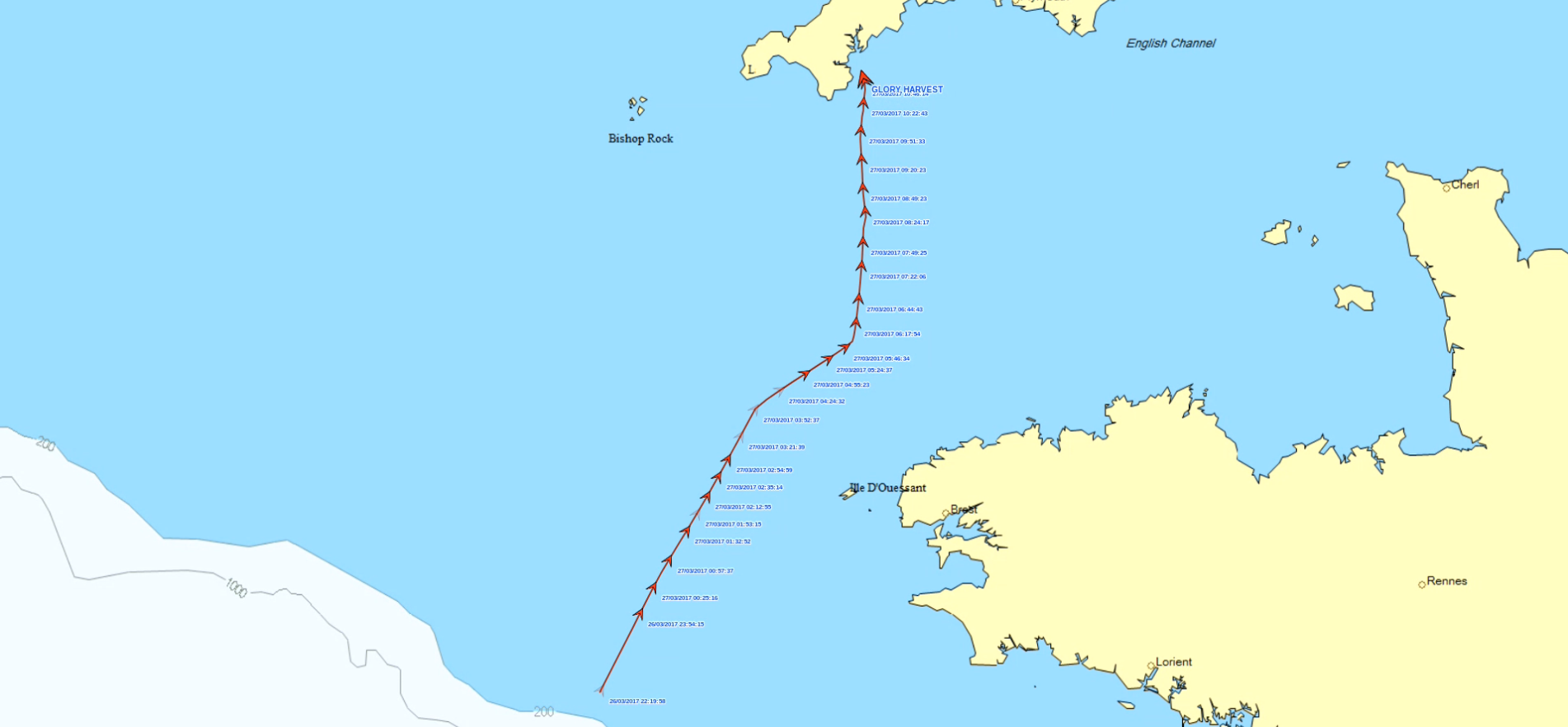}
    \caption{An AIS track flagged as abnormal by the original model but considered as normal by the new model. This track is actually normal. However, in the original training set, there are not many tracks of this type (because they are outside of the coverage zone of the terrestrial AIS station), hence it was flagged as abnormal. The coverage area of the new dataset is better, \GeoTrackNet then succeeded to tag this track as normal.}
    \label{fig:s-ais}
\end{figure}

The combination of S-AIS and T-AIS seems to improve the performance of \textit{GeoTrackNet}. As shown in Fig.~\ref{fig:s-ais}, a vessel which was falsely detected by the original model is correctly flagged as normal by the new model thanks to the improved training dataset.

\newcommand{\eg}{{e.g.,}\xspace}
\newcommand{\GO}{\emph{GeoTrackNet Operator}\xspace}
\newcommand{\GOs}{\emph{GeoTrackNet Operators}\xspace}

\section{Online Detection}
\label{sec:onlineDetection}

Within an operational surveillance system of a given area of interest, it is mandatory
to detect suspicious behaviours as early as possible in order to be able to
react accordingly if necessary. For this application, \GeoTrackNet could be a
tool helping the analysts to sort through the ever increasing flow of AIS data
they have to monitor in the area of interest. In this context, the ability of \textit{GeoTrackNet}, possibly implemented within a distributed framework, to process real AIS data streams and raise alerts in real time is a key issue. 

In this section, we first present the proposed framework for the implementation of  \GeoTrackNet in a streaming context. Then, we report numerical experiments to evaluate scale-up performance. 


\subsection{High level principles}
\label{subsec:highLevelPrinciples}


In a streaming context, \GeoTrackNet is wrapped in a streaming operator: the
\GO. This operator is responsible for (a) preprocessing the stream of AIS
messages and (b) triggering the calls to the anomaly detection
function. The preprocessing phase consists in building incrementally the
tracks from the stream of AIS messages. A track is composed of AIS messages
belonging to the same ship (same MMSI) where erroneous positions (\eg not in
the ROI) or erroneous speed messages (\eg greater than 30 knots) have been
removed. The time difference between consecutive messages in a track is
assumed to be less than a threshold (here 4 hours). Otherwise, they will be
part of two different tracks. Note that tracks are re-sampled to a resolution
of 10 minutes using a linear interpolation.

The \GO must then keep hold of the currently active tracks until a minimum
track duration is reached before triggering the anomaly detection. To cope with a
potentially high velocity of the AIS messages stream, the \GO can be
replicated. In this situation the stream of input messages needs to be
partitioned such that two messages belonging to the same ship are sent to the
same instance of the \GO. We choose to partition the stream of messages
according to the value of the MMSI field to achieve this requirement.

The current implementation of the \GO is based on the
Faust\footnote{https://faust.readthedocs.io} library. This library implements
the Kafka\footnote{https://kafka.apache.org} protocol. The Kafka protocol
allows for partitioning the stream of AIS messages regardless of the number of
\GOs actually deployed. Consequently, the infrastructure can get scaled up and
down when necessary to keep up with the message rate.

\subsection{Synthetic evaluation}

The performance of \GeoTrackNet is critical as it allows both early anomaly
detection and low resource usage. In this section, a quantitative 
evaluation of the computational complexity of \GeoTrackNet is proposed. For this purpose, our benchmarking experiments rely on 
a synthetic set of tracks built from the original dataset~\cite{nguyen_geotracknet-maritime_2019} but covering the 
period from July 2011 to January 2018 \footnote{The code to reproduce the experiment in this Section is available at: https://github.com/msimonin/MultitaskAIS/tree/online\_detection/bench}. 
In total, $237\,863$ tracks were built, $147\,786$ did not pass the preprocessing 
step while $90\,077$ were actually tested for normality. For each tested track the execution 
time of \GeoTrackNet was recorded. Note that this time includes the preprocessing
time. The evaluation has been carried on the Grid'5000 testbed~\cite{balouek_adding_2013}, specifically
on a Dell PowerEdge C6220 of the \textit{Paravance} cluster. 
\GeoTrackNet was given one CPU core and the execution time
statistics are depicted in Fig.~\ref{fig:synthetic}.
\begin{figure}
\begin{center}
    \begin{tabular}{lr}
    mean  & $2.07$ \\
    \hline
    std   & $0.22$ \\
    \hline
    min   & $1.49$ \\
    \hline
    $q_1$ &	$1.93$ \\
    \hline
    median& $2.04$ \\
    \hline
    $q_3$ & $2.19$ \\
    \hline
    max   & $3.74$ \\
    \hline
    \end{tabular}
    \caption{Statistics of the execution time of \GeoTrackNet on $90\,077$
    tracks. On average, it takes $2,07$s to test the normality of a track.
    Only the tracks that passed the preprocessing phase are accounted here.}
    \label{fig:synthetic}
\end{center}
\end{figure}

\begin{figure}
\begin{center}
    \includegraphics[width=\linewidth]{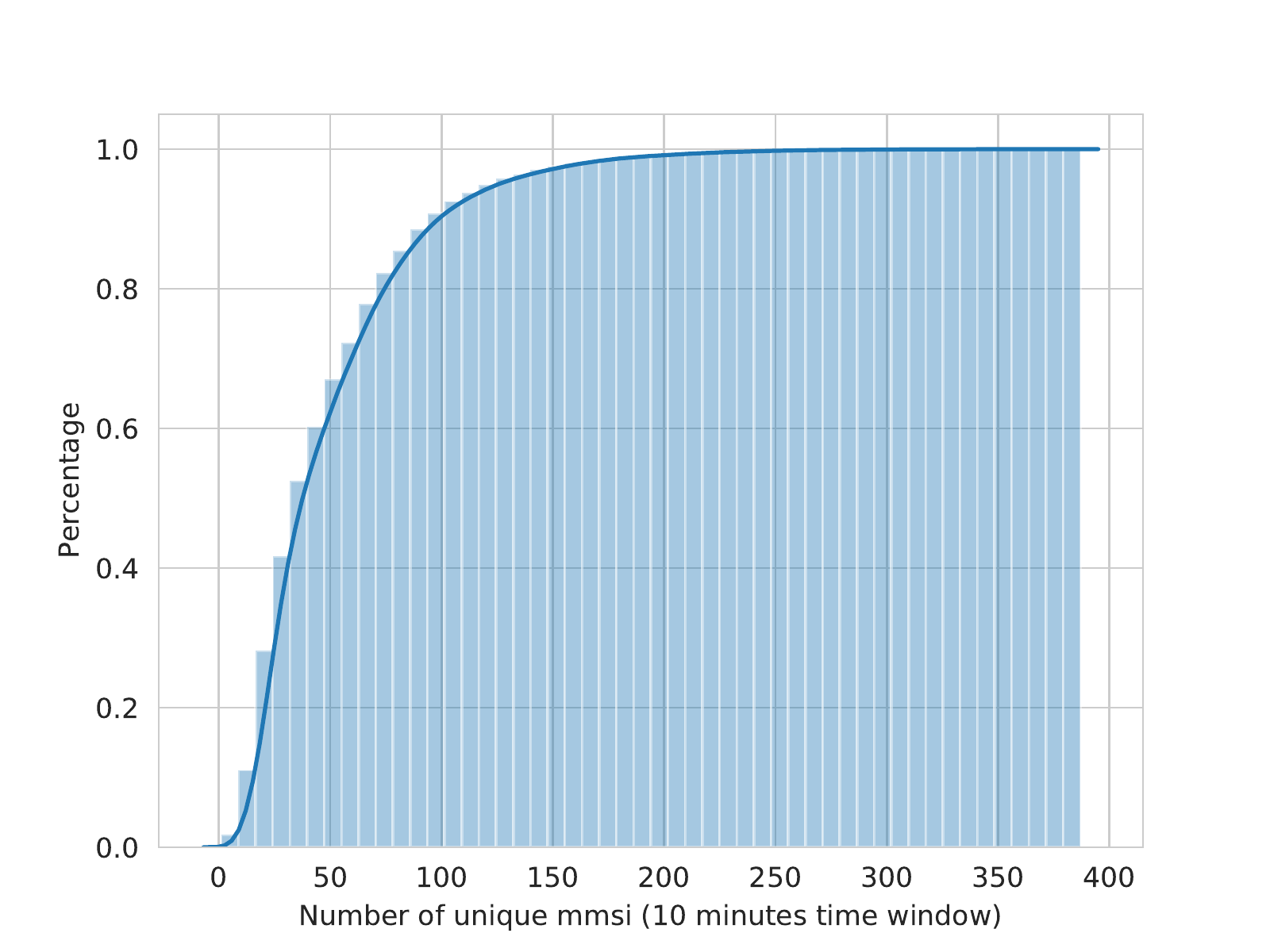}
    \caption{CDF of the number of unique MMSIs received in a 10 minutes time window for the
original dataset (2011-2018). Reading: 80\% of the time less than 80 unique
MMSIs are received on a 10 minutes time window.}
    \label{fig:synthetic:cdf}
\end{center}
\end{figure}
In our setup, \GeoTrackNet can process on average 0.5 tracks per second using one
CPU core. Additionally, due to the preprocessing phase, points are added to a
track every 10 minutes if the rate of messages received for this MMSI is at some point lower than that. Fig.~\ref{fig:synthetic:cdf} depicts the amount of unique MMSI for which at least one message is received in a fixed window. More specifically, according to the graph, a maximum of 400 unique MMSIs are received in a time window of 10 minutes. In other words, for this dataset and ROI, the peak rate at which the \GO has to be called is 400 times per 10 minutes, but most of the time, it is significantly lower that that (as can be seen on the picture, 90\% of the time, less than 100 unique MMSI are received in a time window of 10 minutes). This leads us to the conclusion that 2 CPU cores are sufficient to treat all the tracks of the ROI.

\section{Conclusions and Perspectives}
\label{sec:conclusions}

\GeoTrackNet is a computer-aided system designed to report statistically unusual vessels' tracks. We analysed the validity of this system in terms of relevance of the reported anomalies, scalability of the model, computing performance and possibility to process data streams.

We showed that the model is able to report unusual behaviour that may be difficult to detect with simple rule-based geofencing and limits on course over ground (COG) or speed over ground (SOG). As expected, it fails to detect rare events in area of low traffic. However, most of them can be easily detected by other techniques such as geofencing. It is thus a promising complement in an operational system. 

The current version of \GeoTrackNet does not take into account environmental information. A post-processing step where the ocean current, the wind speed, etc. are used to detect unusual vessels' behaviours due to extreme weather can be added in the future.

\section{Acknowledgment}
\label{sec:Acknowledgment}

This work was supported by public funds (Minist\`ere de l'Education Nationale, de l'Enseignement Sup\'erieur et de la Recherche, FEDER, R\'egion Bretagne, Conseil G\'en\'eral du Finist\`ere, Brest M\'etropole) and by Institut Mines T\'el\'ecom, received in the framework of the VIGISAT program managed by ``Groupement Bretagne T\'el\'ed\'etection'' (BreTel). The authors acknowledge the support of DGA (Direction G\'en\'erale de l'Armement) and ANR (French Agence Nationale de la Recherche) under reference ANR-16-ASTR-0026 (SESAME initiative) and AI Chair OceaniX, the labex Cominlabs, the Brittany Council and the GIS BRETEL (CPER/FEDER framework).

The dataset used in this paper is provided by CLS and Erwan Guegueniat. 

The analysis of abnormal tracks in this paper was carried out using the Maritime Awareness System (MAS) of CLS. 

Experiments in Section \ref{sec:onlineDetection} were carried out using the Grid'5000
testbed, supported by a scientific interest group hosted by Inria and including
CNRS, RENATER and several Universities as well as other organizations.

\bibliographystyle{IEEEtran}
\bibliography{Zotero}
\end{document}